\documentclass[aps,twocolumn,noshowpacs,superscriptaddress]{revtex4-2}

\usepackage{graphicx}
\usepackage{dcolumn}
\usepackage{bm}
\usepackage{mathtools}
\usepackage{xcolor}
\usepackage{layouts}
\usepackage{hyperref}
\usepackage{amssymb}
\usepackage{dsfont}
\usepackage{tabularx}

\usepackage{blindtext}

\newcommand{\tr}{\text{Tr}}
\newcommand{\state}[1]{|#1\rangle}
\newcommand{\conjstate}[1]{\langle #1|}

\newcommand{\1}{\mathds{1}}
\newcommand{\marker}[1]{\raisebox{.5pt}{\textcircled{\raisebox{-.9pt} {#1}}}}
\DeclareMathOperator{\ex}{\mathbb{E}}

\newcommand{\affilITP}{Institute for Theoretical Physics, ETH Z\"{u}rich, CH-8093 Z\"urich, Switzerland.}
\newcommand{\affilKON}{Department of Physics, University of Konstanz, 78464 Konstanz, Germany.}
\newcommand{\affilLAN}{Department of Physics, Lancaster University, Lancaster LA1 4YB, United Kingdom.}

\begin{document}

\preprint{APS/123-QED}

\title{Effect of the readout efficiency of quantum measurement on the system entanglement}

\author{Christian Carisch}\affiliation{\affilITP}
\author{Oded Zilberberg}\affiliation{\affilKON}
\author{Alessandro Romito}\affiliation{\affilLAN}

\date{\today}

\begin{abstract}
Monitored quantum systems evolve along stochastic trajectories correlated with the observer's knowledge of the system's state.
Under such dynamics, certain quantum resources like entanglement may depend on the observer's state of knowledge. 
Here, we quantify the entanglement for a particle on a 1d quantum random walk under inefficient monitoring using a mixed state-entanglement measure -- the configuration coherence.
We find that the system's maximal mean entanglement at the measurement-induced quantum-to-classical crossover is suppressed in different ways by the measurement strength and inefficiency.
In principle, strong measurements can lower the amount of entanglement indefinitely.
However, at a given measurement strength, efficient readout can crucially increase the system entanglement, making high-fidelity detectors essential for successful quantum computing. Our results bear impact for a broad range of fields, ranging from quantum simulation platforms of random walks to questions related to measurement-induced phase transitions.
\end{abstract}

\maketitle

In the noisy intermediate-scale quantum (NISQ) era of quantum computing~\cite{preskill_2018}, quantum measurement has become of crucial practical importance, as it is imperative for harnessing the potential power of quantum devices, as well as for steering quantum systems into desirable states~\cite{wiseman_milburn_2009, harrington_mueller_murch_2022, poepperl_gornyi_gefen_2023}.
Quantum measurements involve a detector performing the measurement and an observer reading out the corresponding measurement outcome.
Due to the coupling of the detector with the system, this process is accompanied by an inherent backaction onto the system~\cite{Born_1926,von_neumann_1996,zilberberg2014measuring,bischoff2015measurement,ferguson_et_al_2023}.
Subsequently, the observer infers the system's state from the measurement record.
Hence, a quantum state reflects the observer's knowledge of the system.
The amount of the observer's gained information on the system's state is determined by the efficiency of the measurement process~\cite{wiseman_milburn_2009,jacobs_2014}.
For example, efficient measurements do not degrade the state-of-knowledge of the observer, while incomplete information about the measurement outcome for inefficient measurements will lead to mixed states of (classical) probabilities associated with possible post-measurement quantum states.
The average over such realizations gives the average system state, which is independent of the readout efficiency and evolves according to a continuous master equation.

Quantum computing relies on harnessing resources encoded in the system's quantum state, like entanglement~\cite{nielsen_chuang_2010}. As entanglement depends non-linearly on the system state, it will also depend on the readout efficiency and, consequently, on the observer's available information. 
One prominent manifestation of such effects is the measurement-induced entanglement phase transition, where a quantum system under fully efficient measurements undergoes a transition from a volume (critical) law to an area law (Zeno) phase with increasing measurement strength~\cite{li_et_al_2018,skinner_et_al_2019,szyniszewski_et_al_2019, chan_et_al_2019, potter_vasseur_2022, fisher_et_al_2023}.
Such measurement-induced transition has been observed in several pioneering experiments~\cite{noel_et_al_2022,koh_et_al_2022, hoke_et_al_2023}.
Strikingly, this transition is strictly absent for the average state, which develops into a complete statistical mixture for any finite measurement strength.
This discrepancy has been analyzed in measurement-induced transitions at finite inefficiency in specific models~\cite{minoguchi_rabl_buchhold_2022, passarelli_et_al_2024}. Recently, the discrepancy was quantified using a mixed-state entanglement measure, where both the fully-efficient and fully-inefficient descriptions feature a measurement-strength dependent coherence length at intermediate times~\cite{carisch_romito_zilberberg_2023}. Apart from these two extreme limits, the dependence of the state's entanglement on the efficiency of the quantum measurement remains to be explored.

In this work, we quantify how the measurement and the observer's knowledge of its outcome impact the system entanglement.
We describe inefficient local density measurements using a stochastic master equation~\cite{jacobs_2014}.
The measurement tends to Zeno-localize the system, whereas inefficient readout leads to mixed-state quantum trajectories.
We analyze these effects using the mixed-state entanglement, which we measure by the recently developed configuration coherence~\cite{van_nieuwenburg_zilberberg_2018, carisch_zilberberg_2023}.
First, we find that for a weakly-measured particle that is coherently oscillating between two sites (the case of charge qubit experiments~\cite{leggett_et_al_1987,gurvitz_et_al_2003,pashkin2009josephson,ferguson_et_al_2023}), the stationary entanglement solely depends on the readout efficiency.
At high measurement strengths, localization drastically suppresses entanglement.
Second, we consider the measurement-induced quantum-to-classical crossover of a particle performing a quantum random walk on a chain, which is a showcase example realized within a diverse range of experimental platforms~\cite{karamlou_et_al_2022,schreiber_et_al_2011,perets_et_al_2008,karski_et_al_2009,schmitz_et_al_2009, zaehringer_et_al_2010,du_et_al_2003, ryan_et_al_2005,broomet_et_al_2010,schreiber_et_al_2010}.
At the quantum-to-classical crossover, the system's mean entanglement assumes a maximal value.
We find that this maximal mean entanglement depends on both the measurement strength as well as the readout efficiency, albeit in quantitatively different ways: the measurement strength suppresses the maximal mean entanglement in a power-law fashion, whereas the readout efficiency increases it exponentially.
Our results demonstrate that the entanglement of a monitored system can increase due to the observer's knowledge of the measurement outcome. 

\begin{figure}[t!]
    \centering
    \includegraphics[width=\columnwidth]{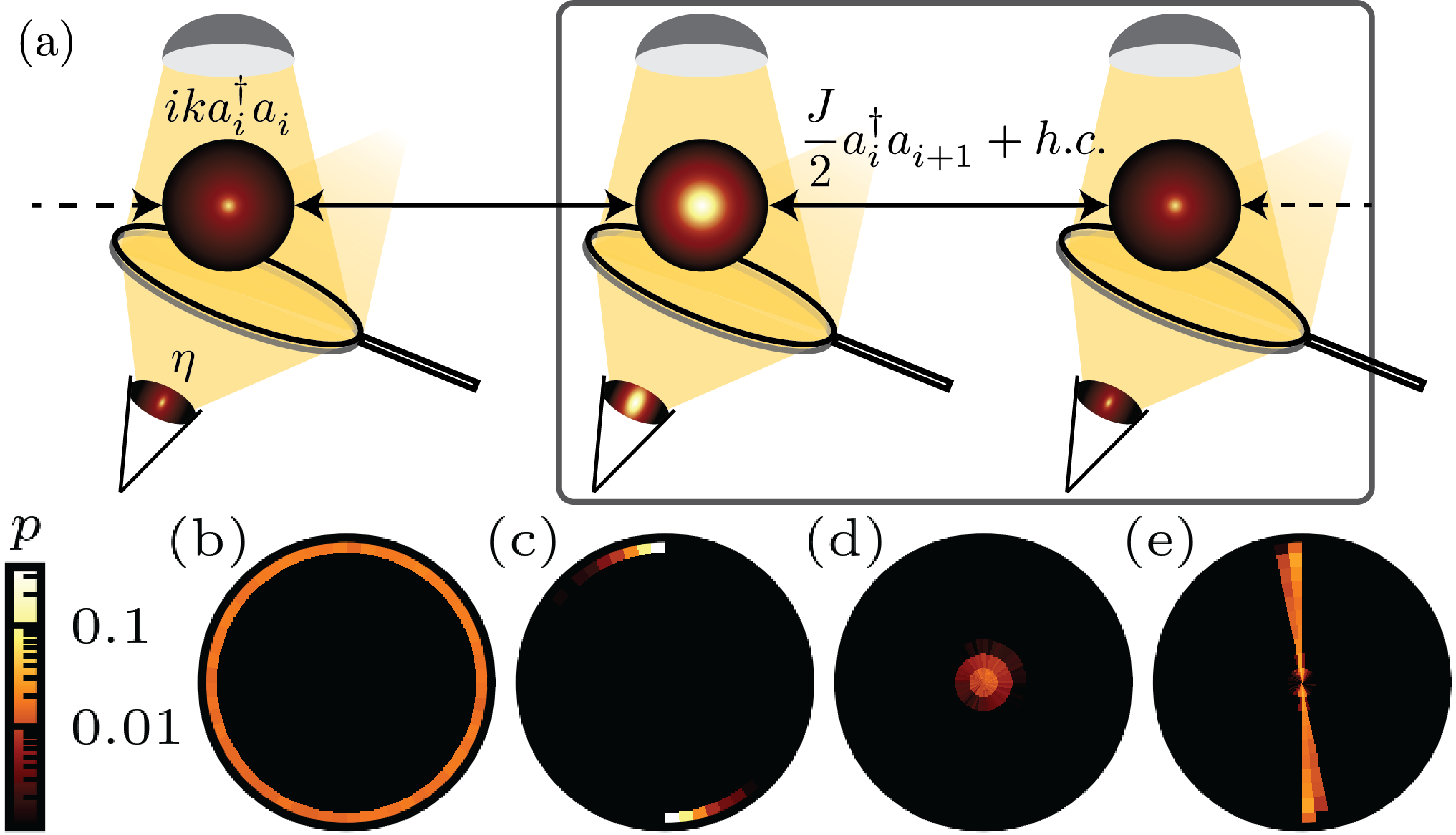}
    \caption{
    (a) Setup: spinless particle hopping on a chain of length $L$ with amplitude $J$. The particle is measured with strength $k$ and efficiency $\eta$ at each site, cf.~Eq.~\eqref{eq: SME}.
    (b-e) Probability densities, $p$, of the stationary state of trajectories for $L=2$ [see rectangle in (a)].
    The Bloch sphere poles correspond to the localized states $|1\rangle$ (north) and $|2\rangle$ (south).
    (b) For high efficiency, $\eta=0.99$, and small measurement strength, $k=0.01$, the trajectories remain almost pure and can cover the full surface of the Bloch sphere.
    (c) For $\eta=0.99$ and high measurement strength, $k=10$, the particle gets projected onto the poles of the Bloch sphere (Zeno effect).
    (d) For $\eta=k=0.01$, the trajectories are highly mixed and distributed symmetrically around the fully mixed state $\1/2$.
    (e) For $\eta=0.01$ and $k=10$, the trajectories get mixed and at the same time projected on the $z$-axis due to the measurement.
    }
    \label{fig: 1}
\end{figure}

We consider a spinless particle hopping on a chain subject to local inefficient density measurements, see Fig.~\ref{fig: 1}(a).
The particle's free evolution follows the Hamiltonian 
\begin{equation}
\label{eq: Hamiltonian}
    H~=\frac{J}{2}\sum_{i=1}^{L-1} a_i^\dag a_{i+1}^{\phantom{\dagger}} + h.c.\, ,
\end{equation}
with $J$ the hopping amplitude, $a_i^\dag$ ($a_i$) the creation (annihilation) operators, and $L$ the chain's length. 
For convenience, we set $J=1$.
We now couple the sites to inefficient detectors that measure the local densities $\langle n_i\rangle = \tr[\rho a_i^\dag a_i], \ i=1,\dots , L$, with strength $k\geq 0$.
The detector efficiency $0\leq \eta \leq 1$ describes the amount of measurement outcome information available to the observer.
With this, the particle's dynamics is described by the stochastic master equation (SME)~\cite{jacobs_2014}
\begin{align}
\label{eq: SME}
    d\rho = -i[H, \rho] dt &+ \sum_{i=1}^L k \left(n_i \rho n_i - \frac{1}{2}\{n_i, \rho\} \right)dt \nonumber \\ 
    &+ \sqrt{\eta k} \left(n_i \rho + \rho n_i-2\langle n_i\rangle \rho \right) dW^i_t \, ,
\end{align}
where the $dW^i_t$ are independent Wiener increments with $ \overline{d W^i_t d W^j_{t'}} =  \delta_{i,j}\delta_{t,t'} d t$.
The second term on the right-hand side in Eq.~\eqref{eq: SME} describes the measurement's backaction on the system, which only depends on the measurement strength.
On the other hand, the third term on the right hand side in the SME characterizes the change of the state due the observer's information gain from the measurement readout.

Due to the randomness underlying the measurement, this term is stochastic and depends on both the measurement strength and the readout efficiency $\eta$.
The evolution of the state $\rho_m$ under a single sequence of measurement outcomes (a single realization of increments $dW^i_t$) is called a quantum trajectory.
Due to $\overline{dW_t^i} = 0$, the ensemble state $\overline{\rho}\equiv (\sum_{m=1}^M\rho_m)/M$ averaged over many trajectories, $M\gg 1$, is independent of the individual measurement outcomes, and evolves according to the standard (continuous) Lindblad master equation~\cite{lindblad_1976, wiseman_milburn_2009}.
Therefore, the average state is independent of the efficiency $\eta$.
An alternative interpretation of Eq.~\eqref{eq: SME} involves two density measurements per site, one with a perfectly efficient detector ($\eta_1=1$) and strength $k_1$, and a second with an inefficient detector ($\eta_2=0$) and strength $k_2$.
This two-measurement picture relates to the inefficient single measurement by $k=k_1+k_2$, $\eta=k_1/(k_1+k_2)$.
Moreover, the inefficient measurement with strength $k_2$ is equivalent to capacitive coupling to a dephasing environment~\cite{jacobs_2014}.

To understand the simultaneous effects of measurement and readout efficiency, we first consider the simple example of two sites, see rectangle selection in Fig.~\ref{fig: 1}(a).
Without measurement, the particle performs Rabi oscillations with frequency $J=1$ between the two sites.
In general, the measurement tends to localize individual trajectories to one of the two sites (quantum Zeno effect~\cite{misra_sudarshan_1977}), whereas inefficient readout will lead to an increasingly mixed state due to the observer's lack of information on the particle's exact location.
To illustrate this effect, we employ a Bloch sphere picture, where the $z$-poles correspond to the localized states $|1\rangle = a_1^\dag |0\rangle$ and $|2\rangle=a_2^\dag|0\rangle$.
For a localized initial state, the $x$-component of the Bloch vector vanishes at all times, and the sphere can be reduced to a disk, see Appendix~\ref{app: sec Bloch sphere}. 

At high efficiencies, the trajectories following Eq.~\eqref{eq: SME} remain almost pure, i.e., on the surface of the Bloch sphere, see Figs.~\ref{fig: 1}(b,c).
This is a consequence of the wavefunction collapse due to the efficient readout: the observer knows which is the specific pure state that corresponds to the measurement outcome with high fidelity.
At small measurement strengths, the Rabi oscillations persist, and at long times the trajectories reach any state on the surface of the Bloch sphere with approximately equal probability [Fig.~\ref{fig: 1}(b)].
Conversely, strong measurements localize the particle on one of the two sites, and the probability distribution at long times is focused at the poles [cf. Fig.~\ref{fig: 1}(c)].
In contrast, for small efficiencies, the restricted amount of the observer's knowledge on the measurement outcomes leads to a mixed state trajectory, see Figs.~\ref{fig: 1}(d,e).
In combination with a small measurement strength, the trajectories end up close to and symmetrically distributed around the maximally mixed state $\1/2 = (\sum_{i=1}^2 |i\rangle \langle i|)/2$ located at the center of the Bloch sphere [cf. Fig.~\ref{fig: 1}(d)].
At high measurement strengths, the individual trajectories' mixing is suppressed by Zeno localization, and the probability distribution skews towards the poles [cf. Fig.~\ref{fig: 1}(e)].
To quantize these effects, we are interested in the entanglement between the two sites.
Interestingly, as we shall see, both the localization and the mixing tend to decrease the amount of entanglement in the system.
Next, we move to discuss how entanglement scales with the measurement strength and the readout efficiency, with consequences on the interpretation of entanglement; we find that it is a resource depending on the amount of information available to the observer rather than a system-inherent quantity.

\begin{figure}[t!]
    \centering
    \includegraphics[width=\columnwidth]{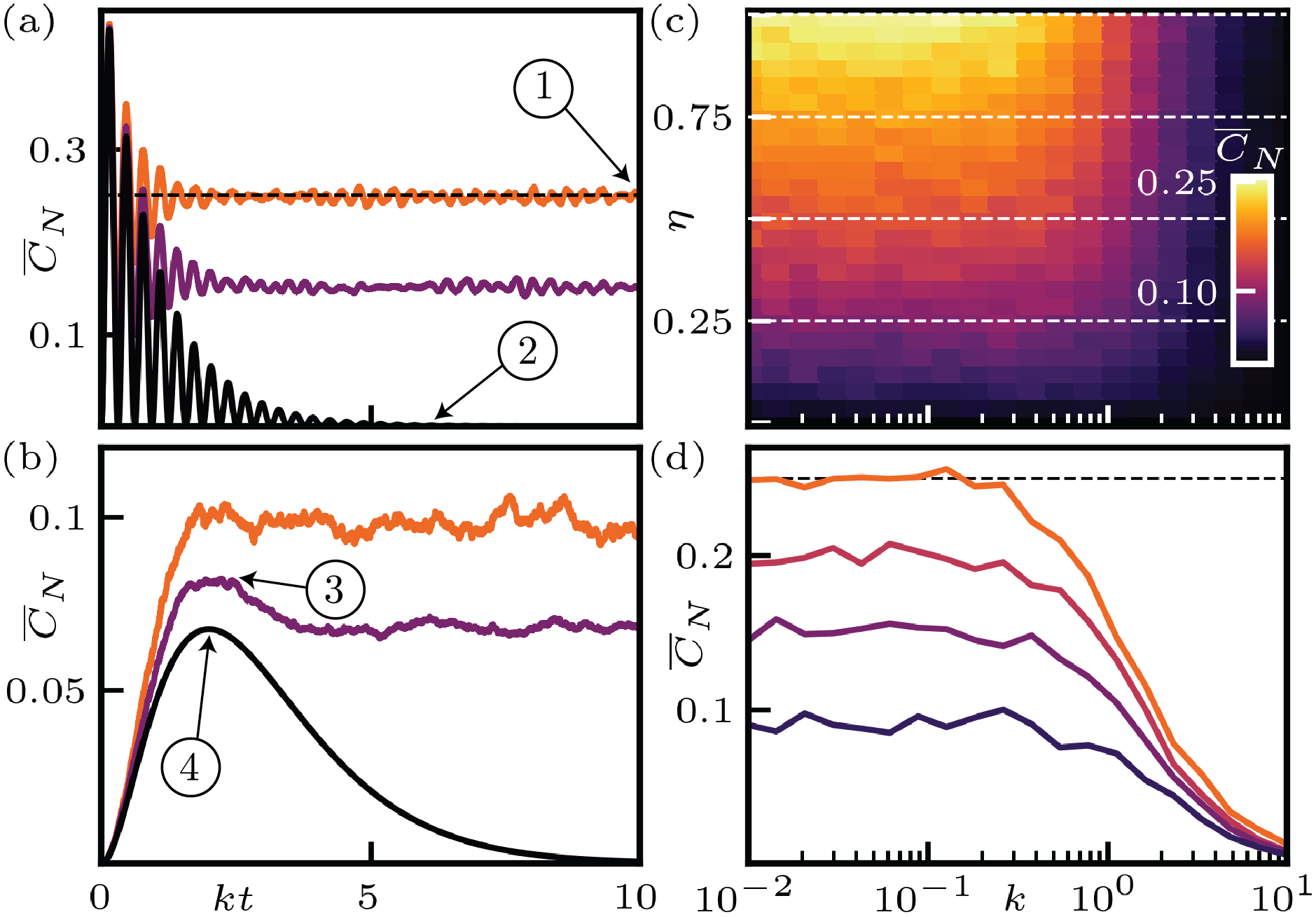}
    \caption{
    (a,b) Time evolution of the configuration coherence for different measurement strengths $k$ and efficiencies $\eta=1, \, 0.5\, , 0$ (top to bottom). 
    (a) For small measurement strength $k=0.1$, the configuration coherence strongly oscillates at first and then saturates at a finite value (for $\eta >0$, marker \marker{1}) or vanishes (for $\eta=0$, marker \marker{2}).
    For $\eta=1$, the stationary value is the average pure state entanglement $\overline{C}_N=1/4$ (dashed line).
    (b) For high measurement strength $k=2$, the entanglement approaches the stationary value in a damped way. For efficiencies $\eta=0, \, 0.5$, there is a maximum at intermediate times (markers \marker{3}, \marker{4}).
    (c) Stationary state configuration coherence as a function of measurement strength and efficiency. Small efficiencies and high measurement strengths suppress the entanglement.
    Horizontal lines highlight the values of the efficiency used in (d).
    (d) Stationary state configuration coherence from (c) for specific values of the efficiency $\eta=1,\, 0.75,\, 0.5,\, 0.25$ (top to bottom). At small measurement strengths, the entanglement only depends on the efficiency, and is equal to the average pure state entanglement $\overline{C}_N=1/4$ for $\eta=1$ (dashed line). At large measurement strengths, the measurement-induced localization suppresses the entanglement.
    }
     \label{fig: 2}
\end{figure}

For a general mixed state $\rho$ of the chain, bipartite entanglement quantifies the quantum correlations across a cut at some bond $1\leq b \leq L-1$.
It can be quantified by the configuration coherence~\cite{van_nieuwenburg_zilberberg_2018, carisch_zilberberg_2023},
\begin{equation}
\label{eq: entanglement}
    C_N(\rho, b) = 2\sum_{\mathclap{\substack{i=1,\dots,b \\ j=b+1,\dots, L }}}|\conjstate{i} \rho \state{j}|^2\, , 
\end{equation}
where $\state{j}=a_j^\dag \state{0}$. 
For systems with a fixed number of particles and Hermitian jump operators such as the SME~\eqref{eq: SME}, the configuration coherence is a convex entanglement measure.
For a single particle, it relates to the negativity $\mathcal{N}(\rho)$~\cite{vidal_werner_2002} as $\mathcal{N}(\rho)=\sqrt{C_N(\rho)/2}$.
For our example of two sites, the configuration coherence reads $C_N~=~2|\langle1|\rho|2\rangle|^2$.
Without measurement, the Rabi oscillations lead to alternating coherent (entangled) and incoherent (non-entangled) configurations of the particle on the two sites.
This persists for small measurement strengths at short times [see Fig.~\ref{fig: 2}(a)] before the entanglement saturates at a stationary value depending on the measurement strength and the efficiency.
As expected, pure Lindblad evolution ($\eta=0$) drives the state to the separable (non-entangled) infinite temperature state $\1/2$.
In conjunction with the persisting Rabi oscillations, the stationary entanglement for $k=0.1$ and $\eta=1$ is $\overline{C}_N\approx 1/4$, which corresponds to the average entanglement on the surface of the Bloch sphere, see Appendix~\ref{app: sec avg pure state ent}.

Whereas the stationary entanglement vanishes for all $k>0$ when $\eta=0$, it remains finite if the observer obtains any information about the measurement outcome [cf. Fig.~\ref{fig: 2}(b)].
As discussed above, both high measurement strengths and low efficiencies decrease the stationary entanglement, the former due to localization and the latter due to mixing, see Fig.~\ref{fig: 2}(c).
Interestingly, for small measurement strengths $k \lesssim 0.3$, the entanglement solely depends on the efficiency, see Fig.~\ref{fig: 2}(d).
Above a critical measurement strength, the measurement succeeds in increasingly localizing the particle.
This can be understood in the following way: the measurement damps the Rabi oscillation and gives it a lifetime. 
When the lifetime becomes shorter than half the Rabi cycle necessary to bring the particle into a coherent superposition between the two sites, the measurement starts to decrease the stationary entanglement.
This observation of an efficiency-only dependent stationary entanglement at small-to-intermediate measurement strengths is a first main result of our work.

\begin{figure}[t!]
    \centering
    \includegraphics[width=\columnwidth]{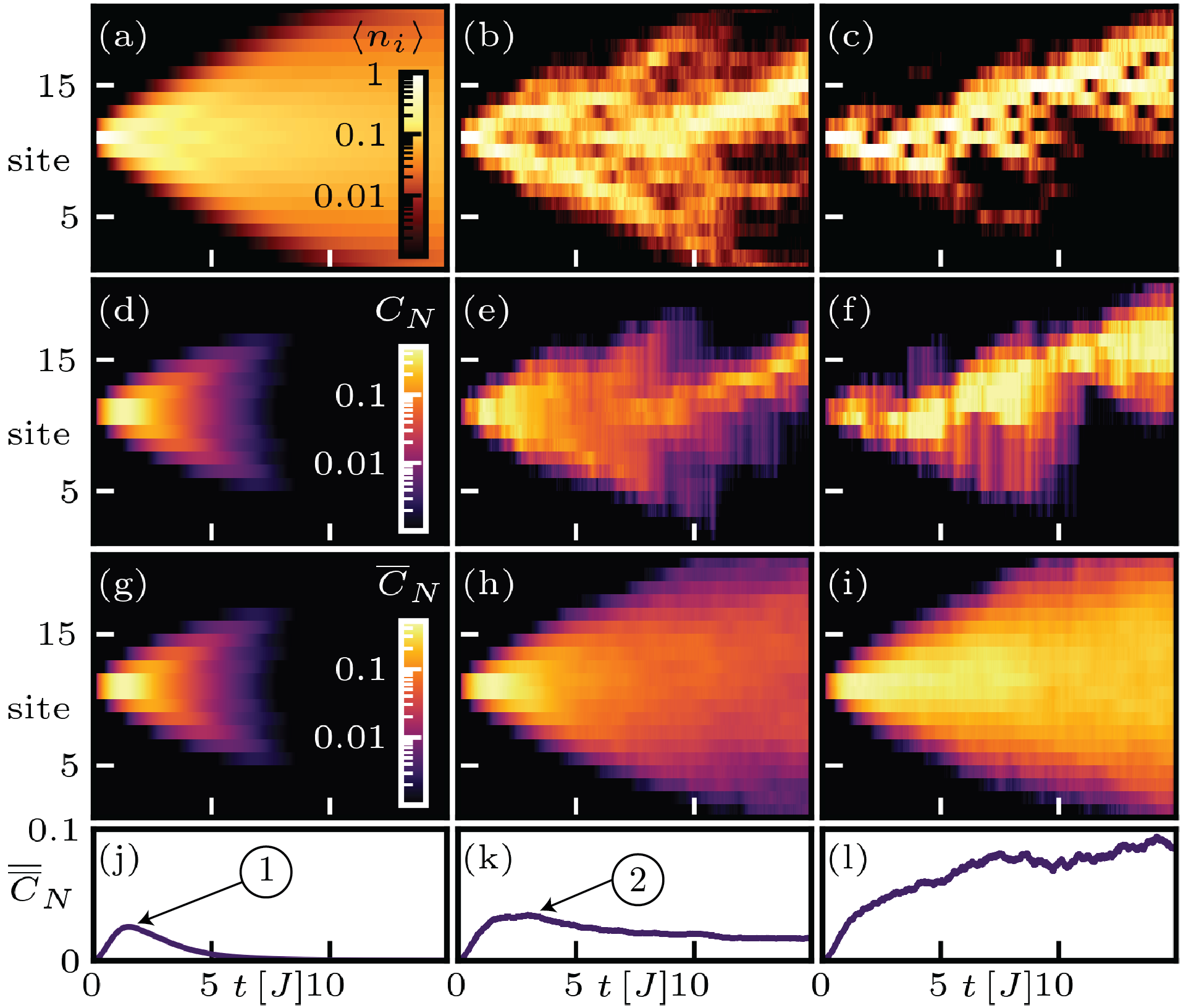}
    \caption{
    (a-c) Local densities $\langle n_i\rangle$ of a quantum trajectory corresponding to a quantum random walk of a single particle injected on site 11 on a chain of length $L=21$, measured with strength $k=0.5$ and efficiencies $\eta=0, \, 0.5, \, 1$ [cf.~Eq.~\eqref{eq: SME}].
    (d-f) Configuration coherence $C_N$ of the random walks in (a-c), respectively.
    (g-i) Configuration coherence $\overline{C}_N$ averaged over many quantum trajectories as in (a-c).
    (j-l) Mean configuration coherence $\overline{\overline{C}}_N$ [cf.~Eq.~\eqref{eq:mean entanglement}] for the evolutions (g-i).
    For $\eta=0$ (first column), the particle undergoes a coherent-to-diffusive crossover. Thereby, the evolution changes from ballistic to diffusive, and the entanglement is absent in the latter. There is no difference between a single trajectory and the average state.
    The mean entanglement has a maximal value at intermediate times (marker \marker{1} in (j)).
    For $\eta=0.5$ (second column) and $\eta=1$ (third column), the trajectories' evolution is interrupted by stochastic jumps.
    In contrast to $\eta=0$, the average configuration coherence in (h,i) shows long-time entanglement.
    For $\eta=0.5$, we observe a maximal mean entanglement (marker \marker{2} in (k)), whereas for $\eta=1$ the maximum is obtained in the stationary limit. 
    }
    \label{fig: 3}
\end{figure}

After having analyzed the effect of measurement and its readout efficiency on the illustrative example of two sites, we consider the single particle hopping on a chain of length $L$.
In that case, the particle performs a quantum random walk~\cite{kempe_2003} governed by the competition between hopping and measurement, see Figs.~\ref{fig: 3}(a-c).
For vanishing efficiency, $\eta=0$, the measurement backaction increasingly mixes the state.
As a consequence, the particle undergoes a quantum-to-classical crossover from a coherent ballistic evolution at short times $t\lesssim 1/k$ to a classical diffusive evolution at long times $t\gtrsim 1/k$~\cite{amir_et_al_2009} [cf. Fig.~\ref{fig: 3}(a)].
With knowledge of the measurement outcomes, $\eta>0$, the quantum random walk becomes interrupted by stochastic jumps whenever the particle is measured at a specific site with high fidelity [cf. Figs.~\ref{fig: 3}(b,c)].
As the crossover is a consequence of the mixing of the state, the evolution between jumps retains its diffusive characteristics for $\eta<1$ [cf. Fig.~\ref{fig: 3}(b)].
At full efficiency $\eta=1$, however, the intermediate evolution is ballistic, see Fig.~\ref{fig: 3}(c).
If the measurement is strong enough, the particle does not have time for coherent spreading between jumps and localizes.
To summarize, the three effects at play act as follows: (i) the hopping tries to coherently spread the particle, (ii) the measurement suppresses such spreading, and (iii) the readout efficiency determines whether the measurement leads to diffusive spreading (low efficiency) or localization (high efficiency).
One natural candidate to analyze such effects is entanglement, which revealed that the particle has a measurement-dependent coherence length for efficiencies $\eta=0, \, 1$~\cite{carisch_romito_zilberberg_2023}.

\begin{figure}[t!]
    \centering
    \includegraphics[width=\columnwidth]{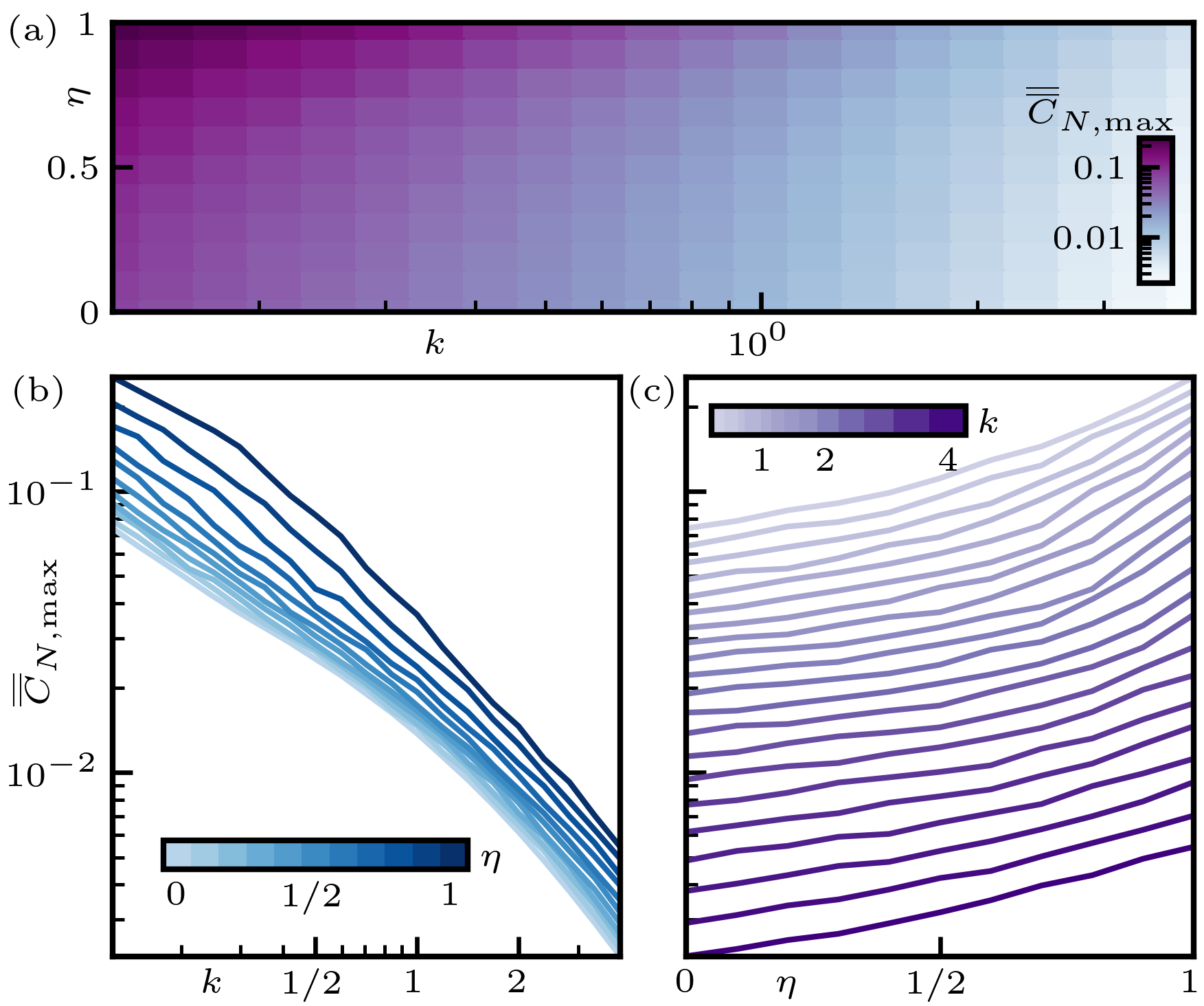}
    \caption{
    (a) Maximal mean entanglement $\overline{\overline{C}}_{N,{\rm max}}$ [cf.~Eq.~\eqref{eq:mean entanglement}] for a chain of length $L=21$ for different values of efficiency $\eta$ and measurement strength $k$ [cf.~Eq.~\eqref{eq: SME}].
    (b) Same as (a) plotted as a function of measurement strength $k$ for different values of readout efficiency $\eta$ using a log-log scale. Straight lines at high measurement strengths $k\geq 1$ hint at power-law scaling, $\overline{\overline{C}}_{N,{\rm max}}\propto 1/k^a$.
    (c) Same as (a) plotted  as a function of readout efficiency $\eta$ for different values of measurement strength $k$ using a semi-log scale. Straight lines at low efficiencies $\eta\leq 1/2$ hint at exponential scaling, $\overline{\overline{C}}_{N,{\rm max}} \propto \exp(b\eta)$. We ascribe deviations from straight lines at small measurement strengths and high efficiencies to finite-size effects.  
    }
    \label{fig: 4}
\end{figure}

Notably, the quantum-to-classical crossover for $\eta=0$ manifests in the entanglement dynamics of the particle: the ballistic evolution is supported by entanglement, whereas in the classical diffusive evolution, entanglement is absent, see Figs.~\ref{fig: 3}(d,g).
Absence of stationary entanglement is expected as the mixed state approaches the separable (non-entangled) infinite temperature state, $(\sum_{i=1}^L|i\rangle\langle i|)/L=\1/L$, for any finite measurement strength $k>0$.
For $\eta>0$, however, the intermediate ballistic evolution between jumps leads to finite long-time entanglement [cf. Figs.~\ref{fig: 3}(e,f,h,i)].
To characterize this effect, we consider the mean entanglement over all bonds
\begin{align}
\label{eq:mean entanglement}
\overline{\overline{C}}_N(t)\equiv \frac{1}{L-1} \sum_{b=1}^{L-1}\overline{C}_N(\rho(t), b)\,.    
\end{align}
Interestingly, at intermediate times, the mean entanglement can assume a maximum $\overline{\overline{C}}_{N,{\rm max}} \equiv \max_t \overline{\overline{C}}_N(t)$ before reaching a stationary value [cf. Fig.~\ref{fig: 3}(j,k)], similar to the single-bond entanglement in Fig.~\ref{fig: 2}(b) for $\eta=0, \, 0.5$.
For small measurement strengths and high readout efficiency, this maximal value is not reached at intermediate times, but in the stationary limit [cf. Fig.~\ref{fig: 3}(l)].
Fig.~\ref{fig: 4} shows the effect of the measurement and the readout efficiency on this maximal mean entanglement.
We find that in general, increasing measurement strength and decreasing readout efficiency suppress the maximal mean entanglement in the system, see Fig.~\ref{fig: 4}(a).
For a given efficiency, the maximal mean entanglement follows a power-law decline with increasing measurement strength, $\overline{\overline{C}}_{N,{\rm max}}\propto 1/k^a$, for some exponent $a$ [cf. Fig.~\ref{fig: 4}(b)].
On the other hand, for a fixed value of the measurement strength, the maximal mean entanglement scales exponentially with the readout efficiency, $\overline{\overline{C}}_{N,{\rm max}} \propto \exp(b\eta)$ [cf. Fig.~\ref{fig: 4}(c)].
However, due to the readout efficiency being bounded, $0\leq \eta \leq 1$, the total entanglement is limited even at maximal efficiency $\eta=1$.
As the measurement strength can be increased indefinitely, $0\leq k < \infty$, the maximal mean entanglement can be lowered below any limit even at perfect readout efficiency.
The differentiated analysis of the effect of quantum measurement and its readout efficiency on the entanglement of the one dimensional quantum random walk is the second main result of our work.

In conclusion, we analyzed the impact of inefficient quantum measurements of a particle hopping on a 1d chain.
To quantify the effect of the measurement and the corresponding readout efficiency, we employed the configuration coherence, a convex mixed state entanglement measure~\cite{carisch_zilberberg_2023}.
For a weakly-monitored particle performing Rabi oscillations between two sites, we find that the stationary entanglement solely depends on the readout efficiency. At high measurement strengths, the measurement-induced quantum Zeno localization of the particle leads to suppressed entanglement.
The quantum Zeno effect in two-level systems is relevant to a broad range of charge qubits~\cite{leggett_et_al_1987,gurvitz_et_al_2003,pashkin2009josephson,ferguson_et_al_2023}, and has also been experimentally observed, e.g. in trapped ions~\cite{itano_et_al_1990}, Bose Einstein condensates~\cite{streed_et_al_2006}, and waveguide arrays~\cite{liu_et_al_2023}.
Also for a particle performing a quantum random walk on a chain of arbitrary length, the maximal amount of entanglement in a system can be arbitrarily limited by using high measurement strengths.
For a fixed measurement strength, however, we find that the observer's knowledge of the measurement outcome exponentially increases the maximal available amount of entanglement.

Our results show that the effect of quantum measurement on the system entanglement depends strongly on the observer's access to the measurement records, over the full range of zero knowledge to perfect readout efficiency.
This has strong implications on how quantum measurement influences the ability of a system to perform tasks demanding entanglement as a resource.
To experimentally analyze such effects, quantum random walks provide an excellent illustration, as they have been realized on a variety of quantum information processing platforms, including superconducting qubits~\cite{karamlou_et_al_2022}, linear~\cite{schreiber_et_al_2011} and nonlinear optics~\cite{perets_et_al_2008}, trapped atoms~\cite{karski_et_al_2009} and ions~\cite{schmitz_et_al_2009, zaehringer_et_al_2010}, nuclear magnetic resonance~\cite{du_et_al_2003, ryan_et_al_2005}, beam splitter arrays~\cite{broomet_et_al_2010}, and fiber loop configurations~\cite{schreiber_et_al_2010}.
In future work, we will address the effect of inefficient measurements for many-body and higher dimensional systems.

\emph{Acknowledgments.} The authors thank A.~Blessing for help with stochastic calculus and acknowledge financial support by ETH Research Grant ETH-51 201-1 and the Deutsche Forschungsgemeinschaft (DFG) - project number 449653034 and through SFB1432, as well as from the Swiss National Science Foundation (SNSF) through the Sinergia Grant No.~CRSII5\_206008/1.

\appendix

\section{Kraus operator evolution scheme for the stochastic master equation}
\label{app: sec Kraus ops}
Here, we explain the numerical procedure used to evolve a state in time according to the stochastic master equation~\eqref{eq: SME}.
To this end, we make use of a more general stochastic master equation with (not necessarily hermitian) jump operators $L_i$, $i=1,\dots, m$ and local inefficiencies $\eta_i$, $i=1,\dots,m$,
\begin{align}
\label{eq: app SME general}
    d\rho &= -i[H, \rho] dt + \sum_{i=1}^m \left(L_i \rho L_i^\dag - \frac{1}{2}\{L_i^\dag L_i, \rho\} \right)dt \nonumber \\ 
    &+ \sqrt{\eta_i} \left(L_i \rho + \rho L_i^\dag-\tr( L_i \rho + \rho L_i^\dag) \right) dW^i_t\, .
\end{align}
The SME~\eqref{eq: SME} follows from~\eqref{eq: app SME general} by taking $L_i = \sqrt{k} n_i$ and $\eta_i = \eta$.
In general, Eq.~\eqref{eq: app SME general} can be integrated using any stochastic differential equation solver, for example, Euler or Euler-Milstein-methods~\cite{kloeden_platen_1992}.
Such methods, however, lead to unphysical states because they do not generally conserve the unit trace and the positivity of the density matrix due to numerical errors.
To guarantee physical states in the evolution, we use the tool of Kraus operators~\cite{nielsen_chuang_2010}.
The Kraus operator update of Eq.~\eqref{eq: app SME general} is given by~\cite{amini_et_al_2011, rouchon_ralph_2015}
\begin{align}
\label{eq: app SME update general}
    \rho + d\rho = \frac{\tilde{M} \rho \tilde{M}^\dag + \sum_{i=1}^m (1-\eta_i) L_i \rho L_i^\dag dt}{\tr\left[\tilde{M} \rho \tilde{M}^\dag + \sum_{i=1}^m (1-\eta_i) L_i \rho L_i^\dag dt\right]} \, ,
\end{align}
with Kraus operator
\begin{align}
\label{eq: Kraus general}
    \tilde{M} &= \1 - \left( iH + \frac{1}{2}\sum_{i=1}^m L_i^\dag L_i\right)dt \nonumber \\
    &+ \sum_{i=1}^m \sqrt{\eta_i} L_i \left(\sqrt{\eta_i} \tr(L_i \rho + \rho L_i^\dag)dt + dW_t^i\right) \nonumber \\
    &+ \sum_{i,j=1}^m \frac{\sqrt{\eta_i\eta_j}}{2} L_i L_j(dW_t^i dW_t^j - \delta_{ij}dt) \, .
\end{align}
This update automatically conserves the unit trace and positivity of the density matrix.
If the time evolution operator $U = \exp(-i H\Delta t/2)$ is available, the update scheme can be improved to~\cite{rouchon_2022}
\begin{align}
    \rho + d\rho = \frac{U\left(M U\rho U^\dag M^\dag + \sum_{i=1}^m (1-\eta_i) L_i U\rho U^\dag L_i^\dag dt\right) U^\dag}{\tr\left[M U\rho U^\dag M^\dag + \sum_{i=1}^m (1-\eta_i) L_i U\rho U^\dag L_i^\dag dt\right]} \, ,
\end{align}
where $M = \tilde{M} + iH dt$.
Inserting $L_i=\sqrt{k}n_i$, $i=1,\dots,L$ and using $n_i n_j=\delta_{ij}n_i$, $n_i^\dag=n_i$, as well as $\sum_{i=1}^L n_i = \1$, we finally find the update scheme
\begin{align}
\label{eq: Kraus update densities}
    \rho + d\rho = \frac{U\left(M U\rho U^\dag M^\dag + (1-\eta) k \sum_{i=1}^L  n_i U\rho U^\dag n_i dt\right) U^\dag}{\tr\left[M U\rho U^\dag M^\dag + (1-\eta) k \sum_{i=1}^L  n_i U\rho U^\dag n_i dt\right]} \, ,
\end{align}
with
\begin{align}
    M &= M^\dag =  \left(1 - \frac{k dt}{2} \right)\1 \nonumber \\
    &+ \sqrt{\eta k}\sum_{i=1}^L  n_i \left(2\sqrt{\eta k} \langle n_i\rangle dt + dW_t^i\right) \nonumber \\
    &+ \frac{\eta k}{2} \sum_{i=1}^L  n_i((dW_t^i)^2 - dt) \,,
\end{align}
where we have introduced the expectation value $\langle n_i\rangle = \tr[\rho n_i]$.

\begin{figure}[t!]
    \centering
    \includegraphics[width=\columnwidth]{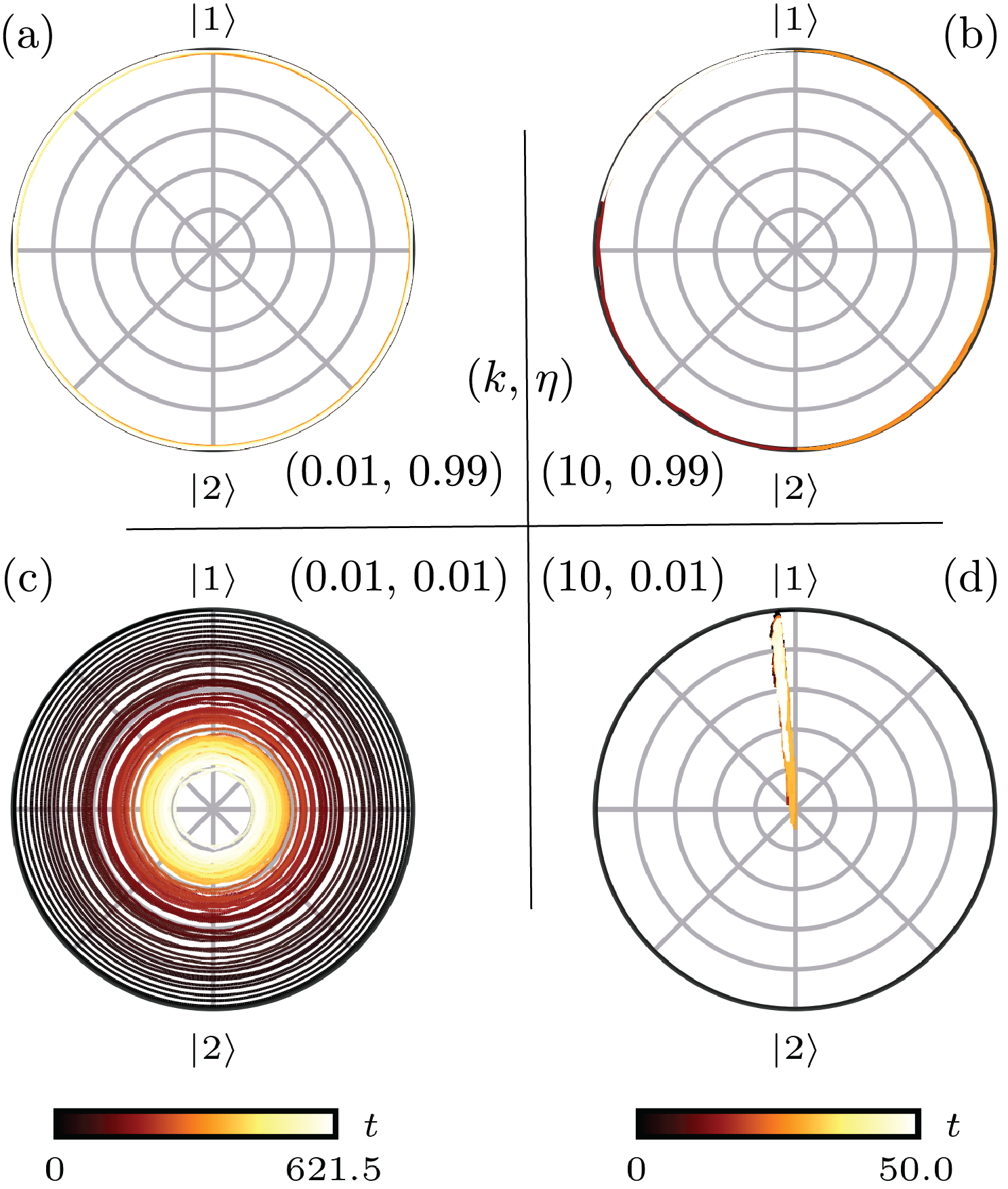}
    \caption{
    Bloch sphere trajectories following Eq.~\eqref{eq: SME} for different values of the measurement strength $k$ and the efficiency $\eta$.
    (a) For small measurement strength and high efficiency, the trajectory performs Bloch oscillations around the surface of the sphere.
    (b) For high measurement strength and high efficiency, the trajectory can still reach all states on the surface of the sphere, but ends up close to one of the poles due to the Zeno effect.
    (c) For small measurement strength and small efficiency, the Bloch oscillations of (a) get damped by the mixing, and the trajectory ends up close to the center of the sphere (corresponding to the maximally mixed state $\1/2$.
    (d) For high measurement strength and small efficiency, the trajectory moves between residing close to the $|1\rangle$ pole and turning into the maximally mixed state due to the competition between the measurement's Zeno effect and the mixing due to the inefficiency.}
     \label{fig: app 1}
\end{figure}

\section{Bloch sphere description of two-site trajectories}
\label{app: sec Bloch sphere}

\begin{figure}[t!]
    \centering
    \includegraphics[width=\columnwidth]{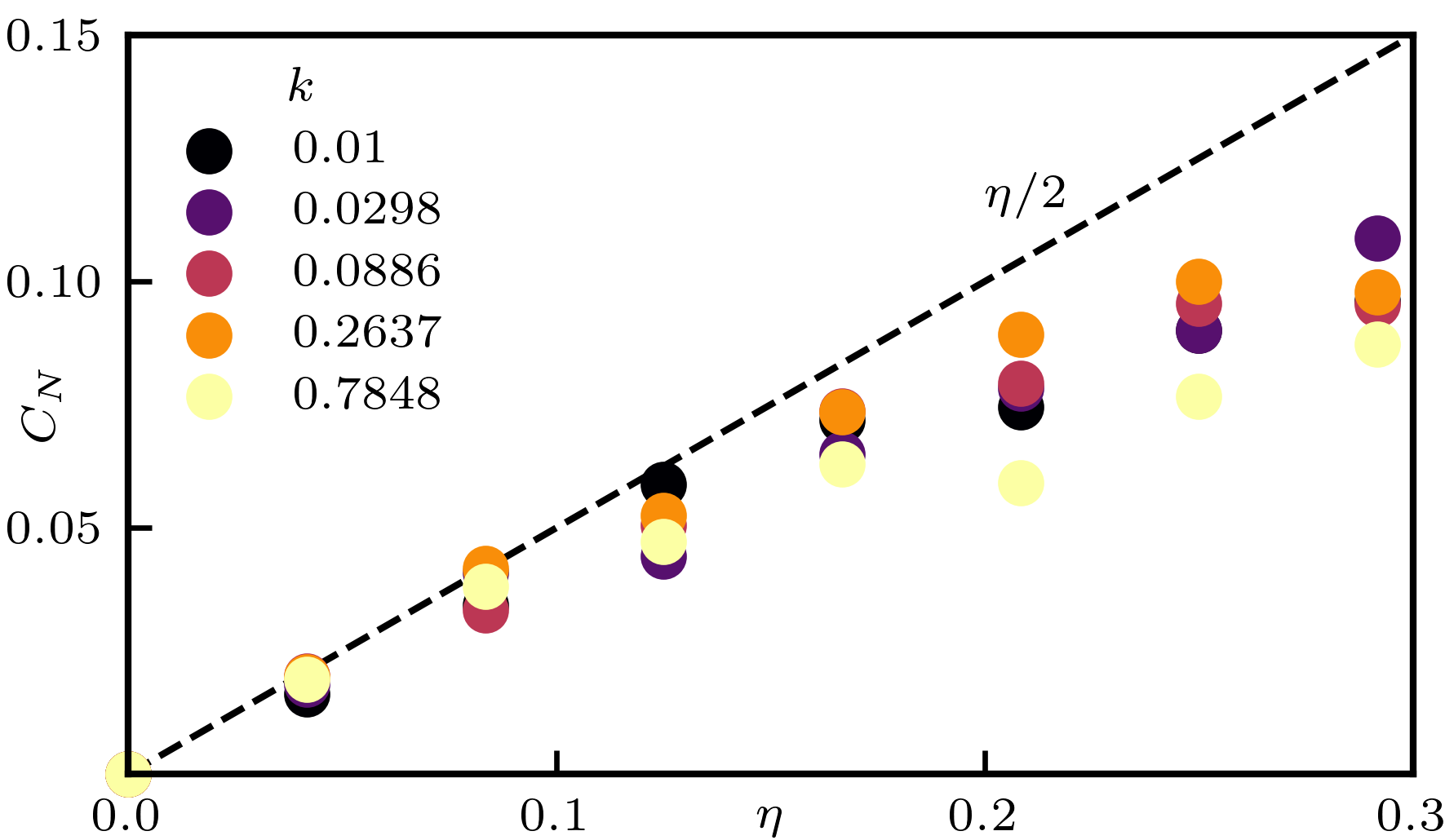}
    \caption{Configuration coherence as a function of the efficiency $\eta$ for several values of the measurement strength $k$.
    For small $\eta$, the scaling approaches the analytical result, $C_N =\eta/2$, cf.~Eq.~\eqref{eq:main result of appendix 2}.}
     \label{fig: app 2}
\end{figure}

In our work, we study the probability distribution of the stationary state of trajectories of a single particle on two sites subject to the SME~\eqref{eq: SME} [cf. Figs.~\ref{fig: 1}(b-e)].
Here, we derive the corresponding Bloch sphere picture and show the dynamics of individual trajectories and how the competing effects of measurement and its efficiency impact their evolution, see Fig.~\ref{fig: app 1}.

First, we note that for a single particle on $L=2$ sites, the Hilbert space is spanned by the site-index states  $(|1\rangle, \, |2\rangle)$ due to the particle number conserving nature of Eq.~\eqref{eq: SME}.
We can then employ a Bloch sphere interpretation of the general 2-site mixed state in the basis $\{|1\rangle, \, |2\rangle\}$,
\begin{align}
\label{eq: Bloch sphere density matrix}
    \rho = \frac{1}{2} \begin{bmatrix}
1+z & x-iy \\
x+iy & 1-z
\end{bmatrix}\, ,
\end{align}
where the vector $\Vec{v} = (x,y,z)$ with $|\Vec{v}|^2=x^2+y^2+z^2=2\tr \rho^2 -1\leq 1$ describes a point on (or inside) the Bloch sphere.

In terms of $x, y, z$, the SME~\eqref{eq: SME} can be written as
\begin{align}
\label{eq: Bloch sphere SDE}
    dx &= - x k dt-xz\sqrt{2\eta k} dW_t  \,,\nonumber \\
    dy &= - yk dt -z dt-yz\sqrt{2\eta k} dW_t \,, \\
    dz &= ydt +(1-z^2)\sqrt{2 \eta k} dW_t \nonumber \, ,
\end{align}
where $dW_t = (dW_t^1 - dW_t^2)/\sqrt{2}$ is a Wiener increment with $\langle dW_t^2\rangle=(\langle (dW_t^1)^2\rangle + \langle (dW_t^2)^2\rangle)/2 = dt$. Eqs.~\eqref{eq: Bloch sphere SDE} describe coupled stochastic differential equations with a quadratic diffusion term. 

Next, we inject the particle in the left site, $\rho(t=0)=|1\rangle \langle 1|$.
For this initial state, we have $x=0$, $\forall t$, and we can restrict the Bloch sphere to the $x=0$ disk.
For small measurement strength and high efficiency, the particle performs Rabi oscillations between the two sites, and remains almost in a pure state, see Fig.~\ref{fig: app 1}(a).
If, on the other hand, the particle is measured strongly, the particle can initially still tunnel between the two sites, but becomes localized close to one of the sites (Zeno effect).
In Fig.~\ref{fig: app 1}(b), the localization happens on the left site.
For small efficiencies, the state becomes more mixed due to lack of information from the measurement readout.
For small measurements, this results in Rabi oscillations towards the center of the Bloch sphere, i.e., the maximally mixed state $\rho = (|1\rangle \langle 1| + |2\rangle \langle 2|)/2$, see Fig.~\ref{fig: app 1}(c).
For high measurements and small efficiencies, the measurement projects the state onto the z-axis of the Bloch sphere, and the small efficiency leads to highly mixed states, as seen in Fig.~\ref{fig: app 1}(d).

For small efficiencies, we can simplify the stochastic differential equations~\eqref{eq: Bloch sphere SDE} in the steady-state regime.
It is known that for $\eta=0$, the steady-state is $\rho=\1/2$, i.e., $x=y=z=0$~\cite{medvedyeva_et_al_2016}.
We can therefore assume that for $\eta\ll 1$, the steady-state is close to the identity and we can neglect the second order terms $\propto z^2$ and $\propto yz$.
As $x=0$, we find the following set of equations,
\begin{align}
    dy &= - yk dt -z dt \,,\\
    dz &= ydt +\sqrt{2 \eta k} dW_t \nonumber \, ,
\end{align}
and the stochastic equations for the quadratic terms read
\begin{align}
\label{eq: quadratic terms}
    dy^2 &= 2ydy+(dy)^2 = - 2y^2 k dt -2 yz dt \,,\nonumber \\
    dz^2 &= 2zdz + (dz)^2 \nonumber \\ &= 2zydt +2z\sqrt{2 \eta k} dW_t +2\eta k dt \,,\\
    d(yz) &= zdy + ydz +dydz = -yzkdt - z^2 dt + y^2dt
    \, , \nonumber
\end{align}
 where we use the rules of It\^o calculus, $dW_t^2=dt$, and neglect all terms of higher order than $dt$.
 
Next, we calculate the expectation values of the quadratic terms $\ex[y^2],\, \ex[z^2]$, and $\ex[yz]$ in the stationary limit, where the left hand sides of the Eqs.~\eqref{eq: quadratic terms} vanish.
Using $\ex[dW_t]=0$, we find
\begin{align}
\label{eq: stationary quadratic terms}
    0 &= - 2\ex[y^2] k -2 \ex[yz] \nonumber \\
    0 &= 2\ex[yz] +2\eta k \\
    0 &= -\ex[yz]k - \ex[z^2] + \ex[y^2]
    \, , \nonumber
\end{align}
with stationary solutions $\ex[yz]=-\eta k$, $\ex[y^2]=\eta$, $\ex[z^2]=\eta (1~-~k)$.
These solutions justify that we can neglect the quadratic terms from the stochastic differential equations for $\eta\ll 1$.
In terms of the trajectory-averaged entanglement, our analysis yields
\begin{align}
\label{eq:main result of appendix 2}
\overline{C}_N=2\overline{|\langle1|\rho|2\rangle|^2}= \ex[y^2]/2=\eta/2\,.
\end{align}
Crucially, we thus find that for small $\eta$, our numerical results from the main text approach this analytical limit independent of the measurement strength $k$, see Fig.~\ref{fig: app 2}.

\section{Average pure state entanglement}
\label{app: sec avg pure state ent}
In our work, we find that for small values of the measurement strength $k$, the long-time entanglement of the single particle on two sites with efficiency $\eta=1$ is $C_N\approx 1/4$.
Here, we show that this value corresponds to the average entanglement of a pure state, in agreement with the fact that for small measurement strengths, the trajectories can explore the whole surface of the Bloch sphere.
Here, too, we can restrict the derivation to the $x=0$-disk if we assume an initial state $|\psi(t=0)\rangle = |1\rangle$, i.e., we can parametrize any pure state on the circle using a single angle $\theta$, $|\psi(\theta)\rangle = \cos(\theta)|1\rangle + i\sin(\theta)|2\rangle$.
This state's configuration coherence is given as $C_N(\theta)=2\cos(\theta)^2\sin(\theta)^2$.
For the average state, we find
\begin{align}
    \overline{C}_N &= \frac{1}{2\pi}\int_0^{2\pi}d\theta 2\cos(\theta)^2\sin(\theta)^2 \nonumber \\
    &= \frac{1}{8\pi}\int_0^{2\pi}d\theta(1-\cos(4\theta)) = \frac{1}{4}\, ,
\end{align}
in conjunction with our numerical findings for small measurement strengths and high efficiency in the main text.

\begin{table*}[ht]
\centering
\begin{tabularx}{\textwidth}{c | c | c | c}
\hline
\hline
Fig. & $dt$ & $t_f$ & \# of trajectories \\
\hline
\ref{fig: app 1} & $\text{min}(1, 1/k)\times 10^{-3}$ & 
$\begin{cases} 
-\log(2 \times 10^{-3} (1-(k/2)^2))/k,\, k<2 \\ 5k,\, \text{otherwise}\end{cases}$ 
& 1 \\
\hline
\ref{fig: 1}(b-e) & $\text{min}(1, 1/k)\times 10^{-3}$ & 
$\begin{cases} 
-\log(2 \times 10^{-3} (1-(k/2)^2))/k,\, k<2 \\ 5k,\, \text{otherwise}\end{cases}$ 
& 10000 \\
\hline
\ref{fig: 2}(a) & $10^{-3}$ & $10/k$ 
& $\begin{cases} 1,\, \eta=0 \\ 1800, \, \eta =0.5 \\ 2200, \, \eta = 1\end{cases}$  \\
\hline
\ref{fig: 2}(b) & $5\times 10^{-4}$ & $10/k$ 
& $\begin{cases} 1,\, \eta=0 \\ 1900, \, \eta =0.5 \\ 1700, \, \eta = 1\end{cases}$  \\
\hline
\ref{fig: 2}(c,d),\, \ref{fig: app 2} &$\text{min}(1, 1/k)\times 10^{-3}$ & $\begin{cases} 
-\log(2 \times 10^{-3} (1-(k/2)^2))/k,\, k<2 \\ 5k,\, \text{otherwise}\end{cases}$ 
& $\text{max}(1000\eta, 100)$  \\
\hline
\ref{fig: 3}(a-f) &$10^{-3}$ & $15$ 
& 1  \\
\hline
\ref{fig: 3}(g-l) &$10^{-3}$ & $15$ 
& 100  \\
\hline
\ref{fig: 4}(a-c) &$\text{min}(1, 1/k)\times 10^{-2}$ & $\text{max}((L/4)^2 k, L)$
& $\begin{cases} 1,\, \eta=0 \\ 200 \lceil 10 \eta k \rceil, \, \text{otherwise}\end{cases}$  \\
\hline
\end{tabularx}
\caption{Details of the numerical implementation leading to the presented results in the figures in the main text and the Appendices.}
\label{table: numerical details}
\end{table*}



\section{Details of the numerical implementation}
\label{app: sec details of numerics}
In Table~\ref{table: numerical details}, we provide numerical details for the simulations leading to the results in the Figures of the main text as well as the appendix.
Our numerical simulations employ the Kraus operator update~\eqref{eq: Kraus update densities} with varying values of the timestep $dt$, stopping times $t_f$, and number of trajectories that are averaged over.
The timestep has to be small compared to the measurement strength, i.e., $k dt\ll 1$, and the number of trajectories that we average over in order to obtain a faithful averaged statement has to be larger for higher efficiency $\eta$, as the stochastic terms in the increment $d\rho$ scale with $\sqrt{\eta k}$.

In order to make statements about the stationary state of average quantities, we run the simulation until the quantities converge.
For the two-site example of Figs.~\ref{fig: app 1} and Figs.~\ref{fig: 1}(b-e), \ref{fig: 2}(c,d) of the main text, we make use of the analytical solution of Eq.~\eqref{eq: SME} at efficiency $\eta=0$ as a reference.
From this solution, the configuration coherence can be calculated analytically~\cite{carisch_romito_zilberberg_2023},
\begin{equation}
\label{eq: analytical C_N}
    C_N(t) = 
    \begin{cases}
        \frac{e^{-k t}}{2\,(1-(k/2)^2)}\sin^2\left(\sqrt{1-(k/2)^2}t\right), \ k<2  \\
        \frac{1}{2}t^2e^{-2t}, \ k = 2 \\
        \frac{e^{-k t}}{2\,((k/2)^2 - 1)}\sinh^2\left(\sqrt{(k/2)^2 - 1}t\right), \ k > 2 \, .
    \end{cases}
\end{equation}
We use this result to estimate a stopping time $t_f$ for our trajectory simulations for $k<2$.
As the average state of the trajectories follows the SME~\eqref{eq: SME} independent of the efficiency $\eta$, the trajectories' stationary state is reached when the average state has reached its steady state, i.e., when the configuration coherence~\eqref{eq: analytical C_N} is very close to zero.
We set as a condition for $t_f$ that $C_N(t_f)\leq 10^{-3}$ and find
\begin{align}
    &\frac{e^{-kt_f}}{2\,(1-(k/2)^2)} \leq C_N(t_f)\leq 10^{-3} \nonumber \\ &\implies t_f = -\log(2 \times 10^{-3} (1-(k/2)^2))/k \, .
    \label{eq: app stopping time}
\end{align}
For measurement strenghts $k\geq 2$, we find good convergence for $t_f = 5k$.

For chain lengths $L>2$, an analytical solution does not exist.
However, if we are interested in the intermediate time dynamics rather than the stationary limit, we can infer a reasonable stopping time $t_f$ from the dynamics.
In the main text, we are interested in the maximal mean entanglement, $\overline{\overline{C}}_{N,{\rm max}}$.
At each bond, the maximal value will be realized shortly after the particle reaches that bond (the entanglement damping due to the measurement will thereafter monotonously decrease the entanglement).
Until the quantum-to-classical crossover at $t\approx 1/k$, the particle evolves ballistically with velocity $J=1$.
For small measurement strengths, the particle will thus hit the wall (and thereby pass the outermost bonds) at $t_f=L/2$.
If the quantum-to-classical crossover happens before the particle reaches the wall, the subsequent evolution is diffusive, with diffusion constant $D=4/k$~\cite{hofmann_drossel_2021}.
Then, the particle reaches the wall at time $t_f = 1/D (L/2)^2 = k (L/4)^2$.
The corresponding numerics result in the main text of a chain with length $L=21$ shows a converged outcome for a stopping time $t_f = \text{max}((L/4)^2 k, L)$ taking into account both types of dynamics. 

\bibliography{library}

\end{document}